# Strontium-84 Enrichments in Presolar Grains Provide First Evidence of *p*-process Nucleosynthesis in Core-collapse Supernovae

Ishita Pal[1], Manavi Jadhav[1,2], Danielle Z. Shulaker[3], Michael R. Savina[3], Marco Pignatari[4,5,6,7], Lorenzo Roberti[4,5,6,8], Heather Bouillion[1], Maria Lugaro[4,5,9,10], Christopher J. Dory[3], Frank Gyngard[11], Noriko Kita[12], and Sachiko Amari[13]

[1] Department of Physics, University of Louisiana at Lafayette, LA 70503, USA; palishita13@gmail.com
[2] International Health Research Institute, Birkirkara, BKR 0937, Malta
[3] Nuclear and Chemical Sciences Division, Lawrence Livermore National Laboratory, CA, USA
[4] Konkoly Observatory, Research Centre for Astronomy and Earth Sciences, HUN-REN, H-1121 Budapest, Konkoly Thege M. út 15-17, Hungary
[5] CSFK, MTA Centre of Excellence, Budapest, Konkoly Thege Miklós út 15-17, H-1121, Hungary
[6] NuGrid Collaboration, http://nugridstars.org
[7] University of Bayreuth, BGI, Universitätsstraße 30, 95447 Bayreuth, Germany
[8] Istituto Nazionale di Fisica Nucleare - Laboratori Nazionali del Sud, Via Santa Sofia 62, Catania, I-95123, Italy
[9] Institute of Physics and Astronomy, ELTE, Eötvös Loránd University, Budapest, Hungary
[10] School of Physics and Astronomy, Monash University, VIC 3800, Australia
[11] Department of Physics and Astronomy, Clemson University, Clemson, SC 29634, USA
[12] Department of Geoscience, University of Wisconsin-Madison, Madison, WI 53706, USA
[13] Department of Physics and McDonnell Center for the Space Sciences, Washington University, St. Louis, MO 63130, USA
*Received 2025 June 27; revised 2025 October 21; accepted 2025 October 23; published 2025 November 17*

## Abstract

This study reports detection of rare *p*-process isotopes within presolar grains. Presolar grains are relic dust grains from dying stars. These microscopic dust particles are found in primitive solar system materials. Their distinct isotopic compositions record the nucleosynthetic processes in their parent stars and the Galactic chemical environment in which these stars formed. We studied presolar graphite grains of high-density type from the Murchison meteorite and found five grains with subgrains that show enrichments in $^{84}$Sr compared to the solar abundance. $^{84}$Sr is the neutron-deficient isotope of strontium that can be produced in the deep oxygen-rich interior of high-mass stars that end their lives as core-collapse supernovae. The observed $^{84}$Sr excesses cannot be produced in low-mass asymptotic giant branch stars, the source of most high-density presolar graphites found in meteorites. High-density graphites with embedded $^{84}$Sr excesses are, instead, compatible with a core-collapse supernovae origin. The graphite subgrains condensed from carbon-rich materials in the outer layers of core-collapse supernovae, where $^{84}$Sr was destroyed by neutron captures during hydrostatic evolution of the stars and their final explosion. Based on current theoretical stellar models, a few percent of contribution from the inner regions of core-collapse supernovae, which are enriched in *p*-process nuclides, to the outer carbon-rich regions is the most likely explanation for the observed enrichment of $^{84}$Sr in the subgrains of the high-density graphites. In this study, we present the first observational evidence that core-collapse supernovae produce and eject isotopes made by the *p*-process.

*Unified Astronomy Thesaurus concepts:* Circumstellar dust (236); Circumstellar grains (239); Stellar Nucleosynthesis (1616); Pre-solar nebulae (1291); P-process (1195); Core-collapse supernovae (304)

## 1. Introduction

Presolar grains are micrometer- to submicrometer-sized dust particles that formed around dying stars before the formation of our solar system. Mineral types of presolar grains include nanodiamonds, silicon carbides (SiC), graphites, oxides, silicates, etc. Each individual presolar grain preserves information about a distinct parent star at a given stage in its evolution (E. Zinner 2014). These grains can be identified by their unique isotopic signatures. The relative abundances of the light element (e.g., C, N, O, Si) isotopes have traditionally helped to characterize the grains and determine the type of stars they originated from (E. Zinner et al. 1995; S. Amari et al. 2001). In fact, measured isotopic compositions of presolar grains can be highly anomalous in some elements (up to 4 orders of magnitude), as compared to the solar or terrestrial values. Such large isotopic variations can only be produced by nuclear reactions activated in the cores of stars. High-precision isotopic measurements (relative to spectroscopic measurements of stars) of multiple elements in presolar grains provide tight constraints to Galactic chemical evolution, stellar evolution, and nucleosynthesis models. Such studies complement spectroscopic measurements of circumstellar dust (E. Zinner 2014).

The evolutionary path of a star, as well as the elements and isotopes it produces, is strongly dependent on its mass. The more massive a star, the heavier the elements (up to Fe) it produces in its deep layers by nuclear fusion involving charged particle reactions (M. Lugaro 2005). Most of the elements heavier than Fe are produced by neutron-capture processes. Depending on the available neutron density, neutron-capture nucleosynthesis can be slow: *s*-process (F. Käppeler et al. 2011); rapid: *r*-process (J. J. Cowan et al. 2021); or intermediate: *i*-process (M. Jadhav et al. 2013a) and the *n*- process (M. Pignatari et al. 2018), relative to the beta decay times of unstable nuclides (E. M. Burbidge et al. 1957). Each of these aforementioned processes produce distinct isotopic abundance patterns.

Currently, we know of up to 35 nuclei heavier than Fe (e.g., $^{84}$Sr, $^{92}$Mo, $^{96}$Ru) that are neutron deficient or proton enriched and can be created by the p-process. The "p-process" is an umbrella term that encompasses various processes that can produce p-enriched nuclides. The most common of these processes is the γ-process or photodisintegration, which can occur in high-energy environments, such as supernovae, where heavier seed nuclei are disintegrated into lighter nuclei via interaction with gamma photons (L. Roberti et al. 2023). However, the main astrophysical site for the origin of the bulk of the p-nuclides remains ambiguous as the γ-process can occur both in Type II core-collapse supernovae (CCSNe) from massive stars (M. Pignatari et al. 2016; C. Travaglio et al. 2018) and in Chandrasekhar-mass thermonuclear (i.e., Type Ia) supernovae (W. M. Howard et al. 1991; C. Travaglio et al. 2011). A significant difficulty in determining the main astrophysical site that produces p-nuclides is that it is not possible to identify p-nuclides from stellar spectra, as their isotopic abundances are a very minor fraction (typically a few percent, with some remarkable exceptions like $^{92}$Mo, which is ∼14.84%) of the total abundances of the elements. The only method that can be used to determine the abundances of such nuclei is the laboratory analysis of physical samples; for example, the analysis of terrestrial samples provides us with the best estimate of the bulk abundances of the p-nuclides in the solar system. Relative to these bulk solar abundances, p-nuclide enrichments have been found so far in various chondritic and achondritic meteorites. Excesses in the p-nuclide of Sr ($^{84}$Sr) have been reported in meteoritic calcium-aluminum-rich inclusions by multiple studies (e.g., F. Moynier et al. 2012; G. A. Brennecka et al. 2013; T. Yokoyama et al. 2015; K. Myojo et al. 2018; Q. R. Shollenberger et al. 2018; C. Burkhardt et al. 2019; B. L. Charlier et al. 2021). One of these studies, B. L. Charlier et al. (2021), attributes the anomalous signatures to presolar phases enriched in p-nuclides within the meteoritic bodies.

Detecting p-process signatures in presolar grains can add significant constraints to determining the astrophysical site(s) of the p-process. To date, a variety of CCSN presolar grains have been identified based on their isotopic compositions but none that unequivocally originated in Type Ia supernovae (SNe Ia; P. Hoppe et al. 2022). Thus, unambiguous detections of p-nuclide excesses in presolar grains with isotopic anomalies compatible with CCSN origins would provide direct evidence that CCSNe are a stellar site of the p-process. Observations of enrichments of p-nuclides in presolar grains have so far been rare and inconclusive. Excesses of p-nuclides $^{124,126}$Xe have been reported in bulk presolar nanodiamonds (R. S. Lewis et al. 1987). However, their origin is still uncertain, and these data represent a bulk analysis of thousands of nanodiamond grains, as nanodiamonds are too small for individual analyses. Excesses in the p-nuclides $^{92,94}$Mo have also been reported in one presolar SiC grain of unclear origin, i.e., belonging to the group of AB grains, which have multiple stellar sources (M. R. Savina et al. 2007). Here, we report the first unambiguous enrichment of the p-nuclide $^{84}$Sr in five high-density (HD) presolar graphite grains, with other measured isotopic abundances that do not rule out a CCSN origin.

Among the three types of carbonaceous presolar phases, presolar graphite grains are the least abundant (∼1–2 ppm) in meteorites, after nanodiamonds and SiC grains. Nevertheless, graphites are on average larger (∼3–5 μm) than most presolar grains, making them easier to analyze as single grains (M. Jadhav et al. 2013b; S. Amari et al. 2014; E. Zinner 2014). They contain tiny refractory subgrains (∼20–500 nm) that may be sources of isotopic anomalies in heavy elements (T. J. Bernatowicz et al. 1991; T. K. Croat et al. 2003, 2005). Presolar graphites are categorized into two density fractions: low-density (LD; 1.6–2 ± 0.1 g cm$^{-3}$) and HD (2 ± 0.1–2.2 ± 0.1 g cm$^{-3}$). A significant portion of the LD graphites have isotopic properties consistent with CCSN ejecta (e.g., S. Amari et al. 1995, 2014; L. R. Nittler et al. 1996; M. Jadhav et al. 2013b). In contrast, HD graphites have multiple stellar sources. The majority of HD grains have isotopic signatures consistent with asymptotic giant branch (AGB) stars, while a minor HD fraction exhibits signatures from CCSNe, post-AGB stars, J-type stars, and novae (M. Jadhav et al. 2013a; P. Haenecour et al. 2016). An important constraint is that due to the facile and dominant formation of CO in cooling circumstellar envelopes, no matter the type of parent star, graphite grains must condense from a chemical environment where C/O > 1 (see the Appendix).

In this work, we present resonant ionization mass spectrometry (RIMS)–measured heavy element isotopic data for five grains that contain distinct $^{84}$Sr anomalies, compared to the solar isotopic value. We present evidence that these anomalies originate in internal subgrains and not in the host grain. Mixing calculations prove that these signatures can be reproduced by mixing different layers from CCSNe. Thus, we conclude that the p-nuclide $^{84}$Sr in the subgrains is synthesized within the interiors of CCSNe and mixed into the ejecta from which the parent grains and subgrains condensed.

## 2. Results

### 2.1. Grains and Subgrains

We studied 49 presolar graphite grains from the HD KFB1 fraction (2.10–2.15 g/cm$^3$) of the Murchison meteorite (S. Amari et al. 1994). The grains were measured for the following stable isotopes of light and heavy elements: C ($^{12,13}$C), N ($^{14,15}$N), O ($^{16,17,18}$O), Sr ($^{84,86,87,88}$Sr), Zr ($^{90,91,92,94,96}$Zr), and Mo ($^{92,94,95,96,97,98,100}$Mo) to understand heavy element nucleosynthesis. Four of the 49 grains were found to have an overall excess in $^{84}$Sr (greater than 2σ), with respect to $^{86}$Sr and solar system values (Figure 1(a)). These four grains had solar values in $^{87}$Sr/$^{86}$Sr and $^{88}$Sr/$^{86}$Sr. All four grains had solar Mo isotopic compositions. Zr isotopic counts in most grains were very low and appeared solar wherever there were enough counts. The C, N, and O isotopic data indicated that the grains with $^{84}$Sr excesses had $^{12}$C/$^{13}$C ratios ranging from 88 to 1777 (solar $^{12}$C/$^{13}$C is ∼89). Their $^{14}$N/$^{15}$N and $^{16}$O/$^{18}$O ratios showed no significant deviation from the solar system and terrestrial values (S. Amari et al. 2018).

While the four graphite grains were found to be overall enhanced in $^{84}$Sr/$^{86}$Sr, the anomalies appeared to be concentrated in certain regions of the grains. In addition to the abovementioned four grains, we noticed enhanced $^{84}$Sr signal within another grain whose overall Sr signature was solar but contained enhanced $^{84}$Sr signal in a certain region of the grain. Our procedure consumed the grains entirely, evaporating material steadily over a period of 10–30 minutes. The Sr signal was generally low during this process but showed occasional bursts, with most of the Sr signal appearing over periods of a few seconds or less. For most grains, the Sr isotopic composition was solar both during the low baseline count rate and the bursts; however, some grains showed enhanced $^{84}$Sr during the bursts. We interpret these bursts as Sr-rich subgrains that carry the p-process anomaly. These are embedded in a graphite matrix in which the Sr concentration is low and the Sr (and Mo and Zr) has solar isotopic composition, likely through contamination either on the parent body or in the laboratory during the chemical procedure used to separate the graphite grains from the host meteorite.

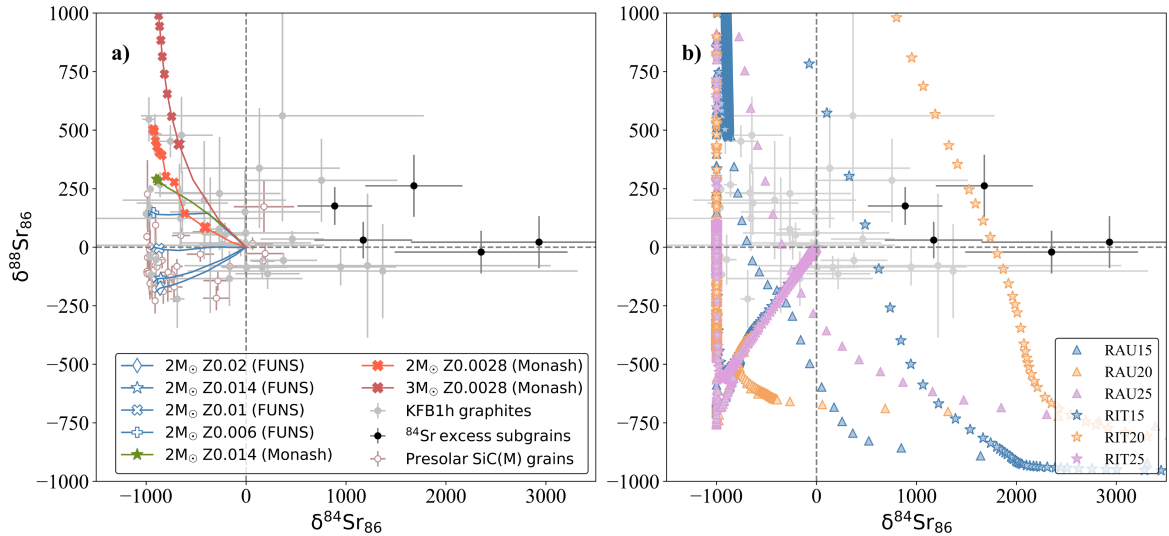

**Figure 1.** Comparison of grain data with stellar models. Three-isotope Sr plots of presolar HD graphites from KFB1 Murchison analyzed in this work. The five subgrains with positive $\delta^{84}Sr_{86}$ values are indicated with black filled circles; all the others are indicated with gray filled circles. All the shown uncertainties represent $1\sigma$ variation. (a) Mainstream SiC grains from the Presolar Grain Database (T. Stephan et al. 2024) are plotted as empty circles. Isotopic ratios of the envelopes in low-metallicity AGB stars (M. Lugaro et al. 2018; D. Vescovi et al. 2020) are shown as blue and green lines as indicated in the legend. Markers along the prediction lines indicate isotopic ratios when the stellar envelope is C rich. (b) Sr isotopic abundance predictions from all interior layers of CCSN models with initial masses of 15, 20, and 25 $M_\odot$ (T. Rauscher et al. 2002; C. Ritter et al. 2018) are plotted in blue, orange, and pink, respectively. See the Appendix for details on AGB and CCSNe models.

Although these subgrains are enriched in one p-nuclide ($^{84}$Sr), they do not show enrichments in the other p-nuclides we analyzed for, $^{92}$Mo and $^{94}$Mo. Molybdenum signals were also concentrated in bursts; however, the Mo and Sr bursts were not correlated, indicating that Mo and Sr are contained in different subgrains inside the parent grains. Graphite grains are known to contain subgrains of TiC and carbides with varying contents of Ti, Mo, Zr, Ru, etc. (T. K. Croat et al. 2003, 2005). It is possible that each of these subgrains condensed from materials from different regions of the progenitor star, with different nucleosynthetic signatures.

In our experience, graphite presolar grains are generally more contaminated with material of solar system isotopic composition than SiC presolar grains, possibly due to the disordered structure of the graphite that permits exchange of intercalated elements in aqueous environments. Nevertheless, we are confident the $^{84}$Sr enhancement observed in the subgrains from this study is not an artifact resulting from contamination in the laboratory or on the parent body. While it is possible that nonresonantly ionized molecular isobars can contribute to the signal at m/z = 84 (e.g., TiC$_3$, Si$_3$, FeSi, and FeCO), these have isotopomers that would be present in our spectra but are not. We further argue against the m/z = 84 signal being artificially enhanced by an isobaric interference because $^{84}$Sr enhancements are always found correlated with Sr-bearing phases, rather than randomly incorporated as one would expect from material infiltrated during the acid dissolution of the meteorite or from aqueous alteration on the parent body.

This analysis is the first heavy element isotopic analysis by RIMS to identify signal bursts consistent with subgrains embedded in the host graphite. As a consequence of the analysis probe size covering the entire parent grain, Sr signals from subgrains were mixed with Sr from the host graphite, thus diluting the anomalies in the subgrains. This means that the $^{84}$Sr anomaly for subgrains is higher than that of the whole grains.

In this Letter, we focus on the five subgrains with excess $^{84}$Sr and the nucleosynthetic origin of their Sr anomalies. The isotopic data of the $^{84}$Sr-enhanced subgrains, along with that of their parent grains, are provided in Table 1. Details of the measurement techniques are provided in the Appendix.

### 2.2. Comparison with Stellar Models

Out of the four Sr isotopes, $^{86}$Sr, $^{87}$Sr, and $^{88}$Sr are predominantly produced by the s-process, while $^{84}$Sr, a proton-rich nuclide, is destroyed by this process (N. Liu et al. 2015). The s-process occurs in the He intershell of low-mass (1.5–4 $M_\odot$) AGB stars, where neutrons are provided by two main sources. The $^{13}$C($\alpha$,n)$^{16}$O reaction is the major contributor of neutrons, with an average neutron density of $10^6$–$10^8$ cm$^{-3}$ activated on timescales of the order of $10^4$ yr. During thermal pulses in stars at the higher end of the AGB mass range, the $^{22}$Ne($\alpha$, n)$^{25}$Mg reaction provides a minor contribution of higher neutron densities (up to $10^{10}$ cm$^{-3}$) for a

Table 1
Light and Heavy Isotopic Data of the Five Presolar Graphite Subgrains with $^{84}$Sr-Excess from KFB1 Murchison

| Isotopic Ratios (‰) | KFB1h-035 | KFB1h-241 | KFB1h-412 | KFB1h-552 | KFB1h-011 |
|---|---|---|---|---|---|
| $\delta^{84}$Sr$_{86}$ | 2932 ± 1282 (1301 ± 631) | 2352 ± 865 (1040 ± 376) | 1158 ± 488 (1064 ± 427) | 1679 ± 486 (693 ± 291) | 887 ± 374 (204 ± 156) |
| $\delta^{87}$Sr$_{86}$ | 52 ± 184 (80 ± 129) | −113 ± 141 (37 ± 79) | 155 ± 126 (95 ± 91) | 130 ± 183 (57 ± 78) | 156 ± 107 (−13 ± 47) |
| $\delta^{88}$Sr$_{86}$ | 22 ± 111 (104 ± 100) | −21 ± 91 (51 ± 49) | −10 ± 78 (21 ± 57) | 262 ± 132 (122 ± 59) | 176 ± 91 (−23 ± 29) |
| $\delta^{92}$Mo$_{96}$ | 126 ± 66 (108 ± 56) | −55 ± 34 (−35 ± 13) | −59 ± 47 (−85 ± 30) | 2 ± 64 (−11 ± 27) | 24 ± 27 (−25 ± 12) |
| $\delta^{94}$Mo$_{96}$ | 171 ± 95 (124 ± 67) | −40 ± 39 (−30 ± 15) | −69 ± 53 (−110 ± 34) | −101 ± 70 (−33 ± 31) | 56 ± 32 (1 ± 14) |
| $\delta^{95}$Mo$_{96}$ | −10 ± 56 (−87 ± 50) | −61 ± 33 (3 ± 13) | −39 ± 46 (−47 ± 29) | 87 ± 65 (11 ± 27) | −16 ± 25 (−38 ± 11) |
| $\delta^{97}$Mo$_{96}$ | 100 ± 69 (17 ± 62) | 15 ± 40 (3 ± 15) | −62 ± 51 (−72 ± 33) | 28 ± 70 (−40 ± 30) | 54 ± 31 (−23 ± 13) |
| $\delta^{98}$Mo$_{96}$ | −127 ± 47 (−106 ± 43) | −43 ± 31 (31 ± 13) | 26 ± 45 (−7 ± 29) | 15 ± 60 (24 ± 26) | −40 ± 23 (−11 ± 11) |
| $\delta^{100}$Mo$_{96}$ | 5 ± 66 (−29 ± 58) | 33 ± 42 (51 ± 16) | −29 ± 55 (−13 ± 36) | −45 ± 72 (−9 ± 32) | 51 ± 32 (10 ± 14) |
| $\delta^{90}$Zr$_{94}$ | −336 ± 1869 (−197 ± 1569) | 3236 ± 12760 (−36 ± 509) | ... | 198 ± 407 (924 ± 815) | −181 ± 94 (−121 ± 70) |
| $\delta^{91}$Zr$_{94}$ | 2993 ± 10713 (2398 ± 7328) | 1769 ± 8613 (−369 ± 440) | ... | −430 ± 442 (−530 ± 424) | 58 ± 38 (49 ± 30) |
| $\delta^{92}$Zr$_{94}$ | 544 ± 4594 (112 ± 2680) | ... | ... | 123 ± 440 (948 ± 852) | −32 ± 34 (−33 ± 27) |
| $\delta^{96}$Zr$_{94}$ | ... | −55 ± 34 (−35 ± 13) | ... | (−74 ± 567) | 108 ± 71 (127 ± 67) |
| $^{12}$C/$^{13}$C | (64 ± 1) | (1033 ± 10) | (90 ± 1) | (88 ± 1) | (88 ± 1) |
| $^{14}$N/$^{15}$N | (280 ± 4) | (277 ± 3) | (272 ± 3) | (275 ± 4) | (272 ± 5) |
| $^{16}$O/$^{18}$O | (460 ± 3) | (467 ± 3) | (457 ± 3) | (443 ± 3) | (462 ± 3) |

**Note.** 1. Whole grain data are given in parentheses. All errors are 1$\sigma$. 2. The C, N, and O ratios are simple ratios, not expressed in units of permil (‰).

period of a few years (F. Käppeler et al. 2011). The s-process also occurs in more massive stars as the $^{22}$Ne($\alpha$, n)$^{25}$Mg reaction is activated during He burning in the core and during C-shell burning (M. Pignatari et al. 2010).

We compare our subgrain data to two AGB stellar nucleosynthesis models (M. Lugaro et al. 2018; D. Vescovi et al. 2020; details in the Appendix), as the majority of HD graphites and the more widely studied mainstream presolar SiC grains (T. Stephan et al. 2024) originate in AGB stars (M. Jadhav et al. 2013b; S. Amari et al. 2014). As expected, these models predict strong $^{84}$Sr deficits (Figure 1(a)). The AGB models by S. Palmerini et al. (2021) predict a very similar trend (personal communications). Stellar envelope compositions during the C-rich phases are particularly depleted in $^{84}$Sr with respect to solar, and they show similar depletions as most mainstream SiC grains (Figure 1(a)). Four mainstream SiC grains plot slightly to the right of the solar reference value, but they have high uncertainties and are disregarded as contaminated (N. Liu et al. 2015; T. Stephan et al. 2018). All the AGB models used in this comparison use an initial solar-scaled abundance composition of Sr. This does not affect comparisons to carbonaceous grains because by the time the AGB star becomes C rich, strong $^{86,87,88}$Sr enrichments by the s-process are also produced, which completely dominate the Sr composition of the envelopes. This is true even if the AGB stars start from a molecular cloud that is already enriched in $^{84}$Sr (see the Appendix). Most of the 49 HD graphite grains we measured are also depleted or solar in $^{84}$Sr, within uncertainty (Figure 1(a)). The five subgrains (within five individual graphite grains) in this study, however, have large and unambiguous $^{84}$Sr enrichments with greater than 2$\sigma$ deviations (see Table 1) from solar values and therefore, require a different stellar nucleosynthesis explanation.

Although a significant percentage of HD grains come from low-metallicity AGB stars (M. Jadhav et al. 2013b; S. Amari et al. 2014), a few appear to originate from other stellar sources. Previous secondary ion mass spectrometry measurements of HD graphite grains reveal excesses in $^{15}$N, $^{18}$O, $^{28}$Si, high $^{26}$Al/$^{27}$Al ratios, and the initial presence of $^{44}$Ti, which unambiguously indicate that the grains originated in CCSNe (M. Jadhav et al. 2013b; S. Amari et al. 2014). Among the different stellar sources proposed as a source of presolar HD graphites, only CCSNe are suitable to host $\gamma$-process nucleosynthesis. In Figure 1(b), we compare the subgrain data with predicted Sr isotopic compositions from models of progenitor stars of initial masses 15, 20, and 25 $M_\odot$ from the Rauscher data set (hereafter labeled as RAU15, RAU20, and RAU25; T. Rauscher et al. 2002) and from the NuGrid data set II (hereafter labeled as RIT15, RIT20, and RIT25; C. Ritter et al. 2018). Compared to AGB stars, the Sr isotopic ratios in CCSN ejecta have a wider range of Sr isotopic values, depending on the mass location of the ejecta. In particular, the inner shell ejecta of RIT15 and RIT20 result in the closest Sr composition to the five subgrains (Figure 1(b)). However, the models do not fit the data perfectly; hence, we invoke mixing of different layers to obtain better fits in Section 2.2.1. More importantly, mixing will allow an adherence to the C/O > 1 constraint as the inner shells of CCSNe are not C rich.

### 2.2.1. Core-collapse Supernovae and Mixing Calculations

While most of the mass of CCSN ejecta is O rich (C/O < 1), the outer zones of the ejecta, where the former convective He-burning shell was located in the stellar progenitor, are typically C rich. Here, some neutron-capture reactions are activated by the $^{22}$Ne($\alpha$,n)$^{25}$Mg reaction during advanced stellar evolution stages. Figure 2 demonstrates the various zones of CCSN ejecta. Zone nomenclature is based on the two most abundant elements in that region of the star (B. S. Meyer et al. 1995). Most of the C-rich outer zones of CCSN ejecta contain s-process nuclides (N. Liu et al. 2015), resulting in an overproduction of $^{86}$Sr and depletion in $^{84}$Sr, i.e., negative $\delta^{84}$Sr$_{86}$.[14] Therefore, materials from the C-rich

---
[14] $\delta^{84}$Sr$_{86}$ (‰) = $\left(\frac{(^{84}\text{Sr}/^{86}\text{Sr})_{\text{grain}}}{(^{84}\text{Sr}/^{86}\text{Sr})_{\text{solar}}} - 1\right) \times 1000$. A general definition of $\delta$ values with standard values used for calculations is available in Appendix A.

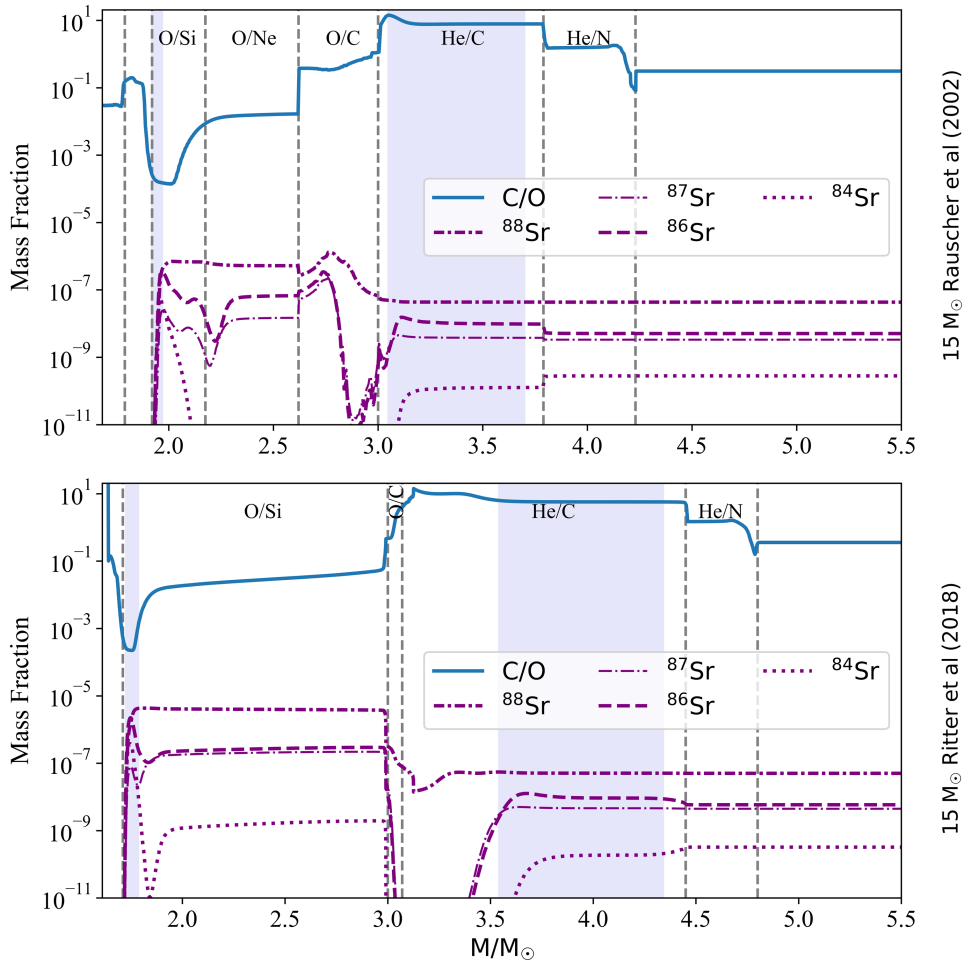

**Figure 2.** Isotopic abundance profiles in CCSNe. Isotopic abundance profiles of 15 $M_\odot$ models from RAU15 (top; T. Rauscher et al. 2002) and RIT15 (bottom; C. Ritter et al. 2018). The isotopic abundances of $^4$He, $^{12}$C, $^{14}$N, $^{16}$O, $^{20}$Ne, and $^{28}$Si are used to assign model layers into zones and are shown in solid thin lines with colors as indicated in the legend. Vertical dashed lines demarcate the stellar zones based on the two most abundant elements in each zone (B. S. Meyer et al. 1995). Isotopic abundances of $^{84,86,87,88}$Sr are plotted with thick purple lines. The two shaded regions in the He/C zones and O/Si zones highlight the regions that can mix to reproduce the observed $^{84}$Sr anomalies, according to our mixing calculations. To reproduce the isotopic patterns of the grains with $^{84}$Sr excesses, we have mixed single layers from the two shaded areas highlighted in the He/C and O/Si zones of the stellar models. Figure 3 shows best-case mixing proportions of these layers for each anomalous subgrain.

He/C zone alone cannot explain the $^{84}$Sr excesses observed in the five HD graphite grains in this study. In contrast, $^{84}$Sr is largely produced in the O-rich innermost regions of the progenitor star via the γ-process (M. Pignatari et al. 2016), especially in the O/Si zones. Hence, mixing of materials between different layers of the CCSNe is required to reproduce the Sr isotopic ratio measured in the grains. Astronomical observations and hydrothermal modeling show evidence for mixing of CCSN ejecta (e.g., J. P. Hughes et al. 2000; N. J. Hammer et al. 2010). Mixing calculations have previously explained anomalies measured in LD graphites from Murchison (e.g., C. Travaglio et al. 1999) and Orgueil (M. Jadhav et al. 2013b), as well as in other presolar phases, such as type-X SiCs (e.g., Y. Lin et al. 2010; T. Stephan et al. 2018) and group 4 oxide grains (e.g., L. R. Nittler et al. 2008). However, the extent and nature of mixing required to explain the isotopic anomalies in the grains are a matter of ongoing debate (M. Pignatari et al. 2013; Y. Xu et al. 2015). Additionally, whether trace elements are incorporated in grains during the condensation process or due to inhomogeneous implantation of ions from deeper regions of the star into the C-rich material during the explosion is not clear (e.g.,

K. K. Marhas & P. Sharda 2018). To explain the anomalous Sr isotopic values of the five subgrains, we model mixing between two separate layers of CCSNe with the mixing fractions taken as free parameters (I. Pal & H. Bouillion 2025). Along with reproducing the grain anomalies, the model also constrains the C/O to greater than 1. According to the model calculations (see the Appendix), mixing layers from 15 $M_\odot$ models (T. Rauscher et al. 2002; C. Ritter et al. 2018) can reproduce the observed isotopic anomalies. As demonstrated in Figure 3, when 0.4%–7.1% of material from an inner layer of the O/Si zone is mixed with 99.6%–92.9% of a layer from the He/C zone of RAU15 and RIT15 (T. Rauscher et al. 2002; C. Ritter et al. 2018), the Sr isotopic ratios measured in the subgrains are well fitted, along with the condition C > O. For RIT20, smaller (0.1%–1.3%) mixing proportions from the inner O/Si layers fit the grain data and fulfill the C/O ratio constraint.

## 3. Discussion

The abundance of $^{84}$Sr is the highest at the base of the O/Si zones as the γ-process efficiency peaks here (Figure 2). Thus,

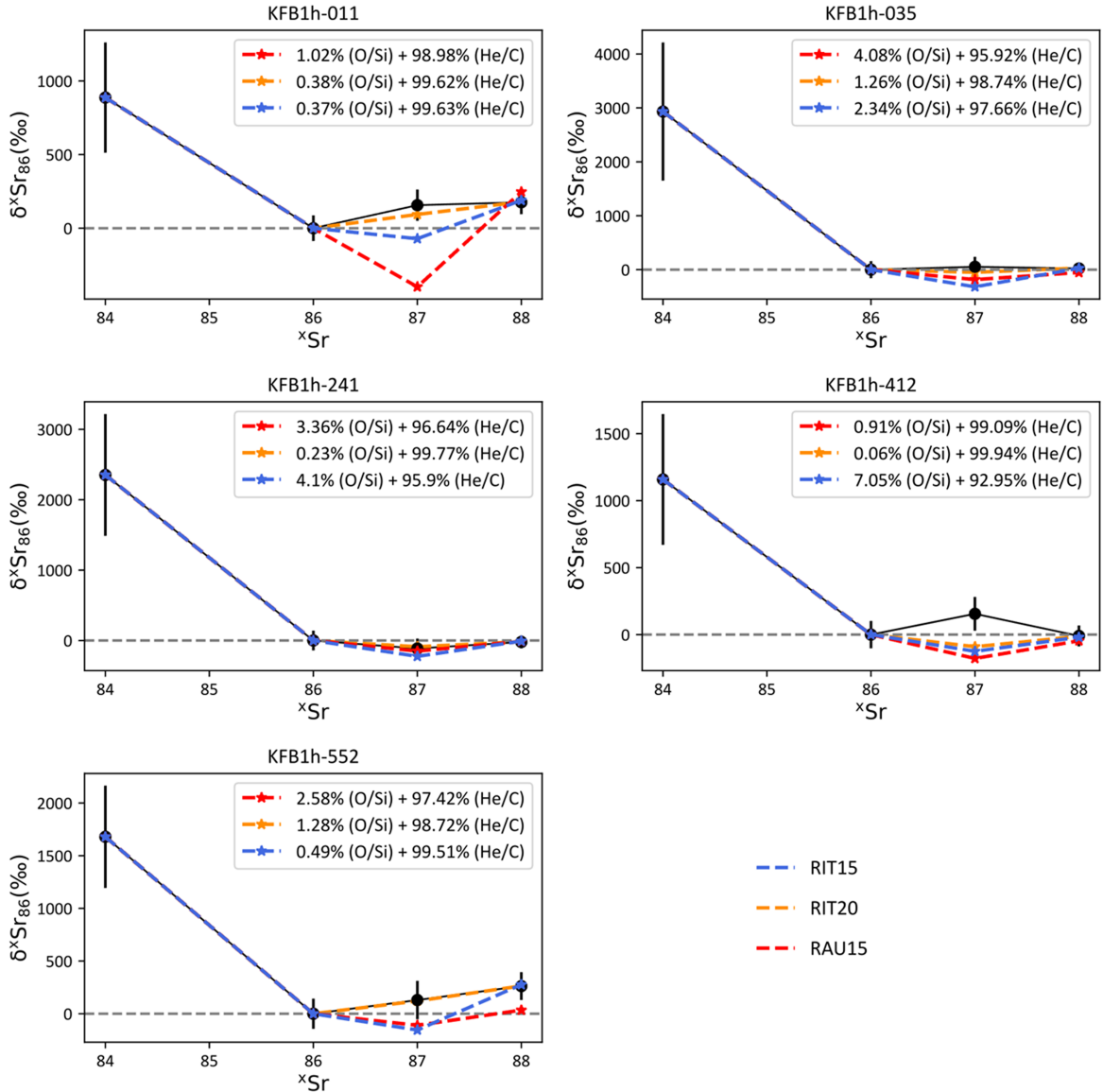

**Figure 3.** Comparison of subgrain data with results from the mixing model. Strontium isotopic ratios versus atomic masses of Sr isotopes for the five subgrains with $^{84}$Sr excesses (black) are shown in each subplot, along with the results from our two-layer mixing model. Uncertainties are 1$\sigma$. We show the best-fit mixing results from three different CCSNe models: RAU15, RIT15, and RIT20 (T. Rauscher et al. 2002; C. Ritter et al. 2018) plotted in red, blue, and orange, respectively. The proportions (in percentages) of the single layers within the specific zones (as defined in Figure 2) that were mixed together are specified for each mixing result. In all cases, the $^{84}$Sr anomalies can be reproduced by mixing small amounts of material from the O/Si zone to the C-rich material of the He/C zone of 15 $M_\odot$ CCSN models by T. Rauscher et al. (2002) and 15 $M_\odot$ and 20 $M_\odot$ models by C. Ritter et al. (2018) See Table A1 for the exact, model-specific, single-layer numbers used in these mixing calculations.

when a minor amount of material from the O/Si zone mixes with the carbon-rich He/C regions of the star, the resulting mixture is enriched in the *p*-process isotope $^{84}$Sr (Figure 3). The mixing model constrains the mixture to be C-rich in order to maintain an environment that is conducive to forming graphite grains in which the anomalous subgrains are found embedded.

We considered two alternative explanations that could also explain the $^{84}$Sr anomalies seen in the grains from this study. The first scenario entails the parent star starting out with an enhanced $^{84}$Sr seed composition compared to solar, and positive $\delta^{84}Sr_{86}$ values are obtained from a mixture of the He-shell region of the star and the normal pristine component of the surface (which was $^{84}$Sr rich compared to solar). Our calculations show that HD graphites with $^{84}$Sr excesses could be used to constrain the initial Sr isotopic abundance of the parent stars of the CCSNe (see the Appendix). Therefore, although we cannot rule out this hypothesis by comparison with stellar models, we find it unlikely based on current observational data (details in the Appendix). In the second scenario, we discuss the possibilities of thermonuclear supernovae as the progenitors of the *p*-process enriched grains in this study. We dismiss this hypothesis as well, based on current observational data and theoretical studies. The details of these scenarios are provided in the Appendix. Thus, we identify the $\gamma$-process nucleosynthesis signature from parent CCSNe as the most viable explanation for the $^{84}$Sr enrichments measured in the subgrains from this study. Further studies of heavy element isotopic systems in presolar grains will provide additional context to assess these models.

We report five presolar subgrains within HD graphite grains that contain statistically significant excesses in $^{84}$Sr; the *p*-only isotope of Sr. AGB stars cannot be the stellar source of these grains because the *s*-process is the dominant nucleosynthesis process in such stars and AGB stars are expected to produce dust grains and subgrains that are depleted in $^{84}$Sr. Instead, the reported grains in this study are most likely CCSN condensates. These data provide the first direct evidence that *p*-nuclides are produced and ejected by CCSNe. The mixing of nucleosynthesis products mostly from the C-rich He shell with minor amounts of material from the inner regions of pre-supernova high-mass stars results in condensates that replicate the Sr isotopic values measured in the subgrains presented in this study.

### Acknowledgments

We thank Tom Yuzvinsky and Yang Mu for electron microscopy and ion milling work at the W. M. Keck Center for Nanoscale Optofluidics at UCSC and the Shared Instrumentation Facility at LSU. I.P. acknowledges NASA FINESST award 80NSSC22K1330. M.J. acknowledges NSF 20–543 award #2033380. M.R.S. and D.Z.S. acknowledge funding from the LLNL Glen T. Seaborg Institute and LLNL-LDRD Program under Project 20-ERD-030. M.R.S. and D.Z.S. work was performed under the auspices of the U.S. Department of Energy by Lawrence Livermore National Laboratory under Contract DE-AC52-07NA27344. S.A. acknowledges support by grant 80NSSC22K0360 from the NASA Emerging Worlds program. N.K. acknowledges partial support of WiscSIMS by NSF (EAR1658823). M.P., L.R., and M.L. thank the support from the NKFI via K-project 138031. L.R. acknowledges the support from the ChETEC-INFRA—Transnational Access (Project 22102724-ST) and access to "viper," the University of Hull HPC Facility. M.P. and L.R. acknowledge the support to NuGrid from the National Science Foundation (NSF, USA) under grant No. PHY-1430152 (JINA Center for the Evolution of the Elements). M.P., M.L., and L.R. thank the Lendület Program LP2023-10/2023 of the Hungarian Academy of Sciences, the ERC Consolidator Grant funding scheme (Project RADIOSTAR, G.A. n. 724560, Hungary), the ChETEC COST Action (CA16117), supported by the European Cooperation in Science and Technology, and OISE-1927130: The International Research Network for Nuclear Astrophysics (IReNA), awarded by the US National Science Foundation. M.P., M.L., and L.R. acknowledge support from the ChETEC-INFRA project funded by the European Union's Horizon 2020 Research and Innovation program (grant Agreement No. 101008324). M.P. also acknowledges the support from the ERC Synergy Grant Programme (Geoastronomy, grant agreement No. 101166936, Germany). M. L. was also supported by the NKFIH excellence grant TKP2021-NKTA-64. L.R. acknowledges the support from the ChETEC-INFRA—Transnational Access Projects 22102724-ST and 23103142-ST and the PRIN URKA grant No. 2022RJLWHN.

## Appendix A
## Methods

### A.1. Nano-Secondary Ionization Mass Spectrometry

Fifty-six individual grains from the KFB1 high-density fraction (2.10–2.15 g cm$^{-3}$) of Murchison were mounted on a gold foil and documented for locations and sizes using a scanning electron microscope (SEM) JEOL JSM-840A at Washington University, St. Louis. Isotopes of C and N were measured using the Nano-Secondary Ionization Mass Spectrometry 50 (NanoSIMS50) at Washington University. Synthetic SiC grains were used as standards. Counts on $^{12}$C$^-$, $^{13}$C$^-$, $^{12}$C$^{14}$N$^-$, and $^{12}$C$^{15}$N$^-$ were collected simultaneously (S. Amari et al. 2018).

### A.2. Secondary Ionization Mass Spectrometry

Oxygen isotope ratios were measured on the same grains using the CAMECA IMS 1280 at the University of Wisconsin-Madison. During the analysis, a focused Cs$^+$ beam of 5 pA was used, and secondary ions of $^{16}$O$^-$, $^{17}$O$^-$, and $^{18}$O$^-$ were detected simultaneously using electron multipliers. Oxygen isotopic ratios were corrected using UWMA1 standard (K. H. Williford et al. 2016; A. Ishida et al. 2018), which is terrestrial organic matter with ~2 wt% oxygen (S. Amari et al. 1994; R. Tartèse et al. 2017).

### A.3. Resonance Ionization Mass Spectrometry

Following light element analysis, the grains were focused ion beam (FIB)–welded to the gold mount and etched with fiducial marks for heavy element analysis. After the NanoSIMS and Secondary Ionization Mass Spectrometry (SIMS) measurements, only 49 of the total 56 graphite grains had enough material left for reliable RIMS analysis. FIB and mount redocumentation work were done at the Shared Instrumentation Facility at Louisiana State University, the W. M. Keck Center for Nanoscale Optofluidics at the University of California Santa Cruz, and the SEM facility at the Lawrence Livermore National Laboratory (LLNL). While some grain material is sputtered away during SIMS work, a substantial amount of grain material remained for the RIMS analysis (see Figure A1). Based on the initial SEM observations of mineralogy and morphology, the grains containing the $^{84}$Sr-excess subgrains did not display any noticeable variations from the rest of the KFB1h grains.

The 49 graphite grains from the KFB1 density fraction of Murchison were measured for isotopes of heavy elements Sr, Mo, and Zr, using the Laser Ionization of Neutrals (LION) RIMS facility at LLNL. The RIMS technique uses tunable lasers to selectively ionize specific elements of interest, which effectively eliminates isobaric interferences. The new-generation RIMS instruments have a higher ion yield compared to SIMS and previous generation RIMS instruments, making them ideal for studying trace element isotopes in presolar grains (M. R. Savina et al. 2003, 2018). Four isotopes of Sr ($^{84,86,87,88}$Sr), five isotopes of Zr ($^{90,91,92,94,96}$Zr), and seven isotopes of Mo ($^{92,94,95,96,97,98,100}$Mo) were simultaneously measured on each grain and Mo$_2$C, ZrC, and SrTiO$_3$ standards with LION. First, a 1–2 $\mu$m spot on the sample grains was volatilized using an Nd:YAG laser (1064 nm, 1000 Hz, 7 ns full width at half maximum). The neutral atoms were then resonantly ionized for all 16 isotopes by six tunable Ti: Sapphire lasers. The lasers were collinearly arranged 1 mm above the sample surface employing a two-color two-photon resonance ionization scheme for each element (J. G. Barzyk et al. 2007; M. Lugaro et al. 2023). The photoions were then accelerated into a time-of-flight mass spectrometer. Isobaric interferences between Mo and Zr isotopes were avoided by slightly delaying the laser pulse for Zr. The ionization lasers for Sr and Mo were first pulsed on the neutral cloud of volatilized grain material, while the sample potential was raised to 3 kV to immediately accelerate the Sr and Mo ions to the mass spectrometer. After a 200 ns delay, Zr ionization lasers were fired on the same neutral cloud, followed by acceleration to the spectrometer. This difference in the time of birth of ions was sufficient to resolve the normally isobaric peaks at $^{92,94,96}$Mo and $^{92,94,96}$Zr.

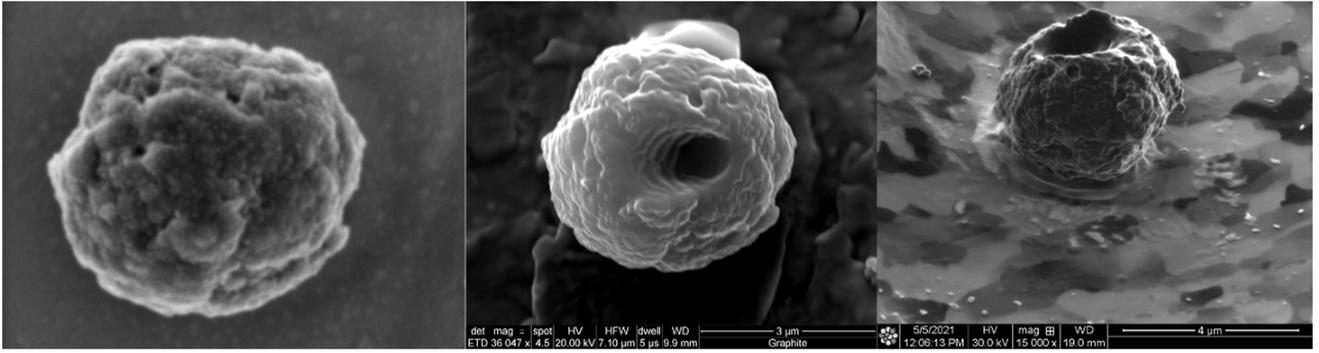

**Figure A1.** Images of grain KFB1h-011 before any SIMS analysis (left) and after SIMS analysis (middle and right). The example images provide ample evidence of there being enough grain material left over for RIMS analysis after SIMS analysis

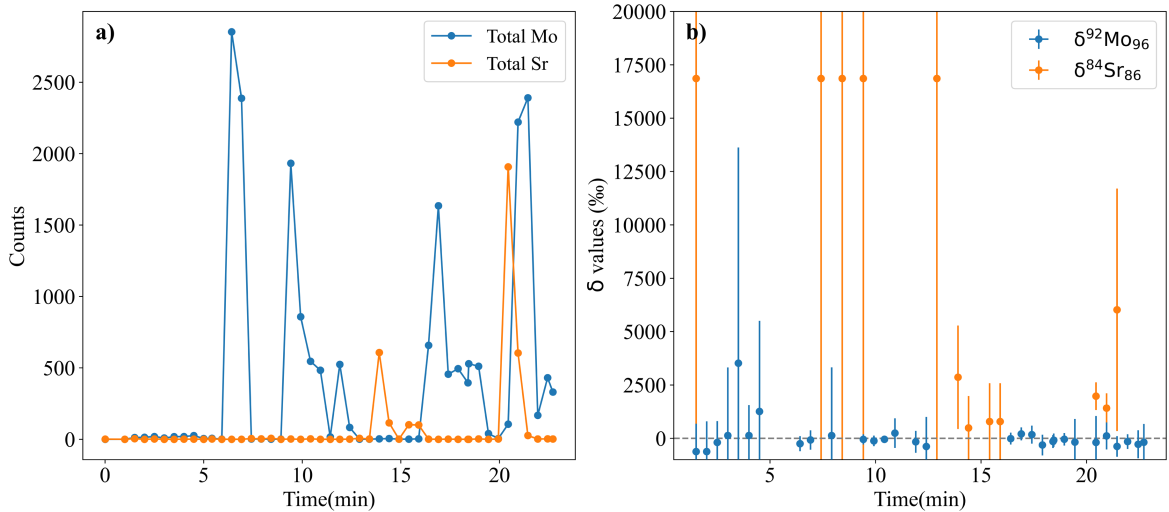

**Figure A2.** Time evolution of the RIMS signal for grain KFB1h-412. (a) The variation in Sr and Mo elemental counts in 30 s time bins, corresponding to 30,000 laser shots each. (b) The $\delta(^{84}\text{Sr}/^{86}\text{Sr})$ and $\delta(^{92}\text{Mo}/^{96}\text{Mo})$ values for each bin. Large error bars correspond to very few counts of $^{84}$Sr.

### A.4. Data Reduction

The data were reduced by normalizing to solar standards and are expressed as $\delta$-values (part-per-thousand deviation from standard ratios):

$$\delta^A\text{Sr}_B = 1000 \times (R_{\text{sample}}/R_{\text{standard}} - 1), \quad (A1)$$

where $R_{\text{sample}}$ is the A:B isotope ratio measured in the grain and $R_{\text{standard}}$ is the same isotope ratio measured in the standard. Strontium solar isotope abundances are $^{84}$Sr = 0.0056; $^{86}$Sr = 0.0986; $^{87}$Sr = 0.0700; and $^{88}$Sr = 0.8258 (J. Meija et al. 2016).

Strontium-84 counts were very low in most of the grains because $^{84}$Sr represents less than 1% of the total abundance of Sr. In our data reporting, we excluded grains in which the $^{84}$Sr counts were consistent with zero at $2\sigma$ (counting statistics including background correction). Thirty-five of the 49 grains yielded enough counts to calculate significant delta values for $^{84}$Sr. Excesses in $^{84}$Sr were identified in regions of five different grains, four of which showed overall (whole grain) positive delta values. After careful consideration, we ruled out interferences from polyatomic C at mass 84 and fragments of organic molecules or isotopomers of inorganic polyatomic species in the vicinity of mass 84. Peak integrals were corrected by integrating and subtracting counts from the adjacent regions before and after the mass peaks. In the rare cases where the noise was highly asymmetric, we used the higher side as our background signal, though this affected only three grains, none of which had enhanced $^{84}$Sr. For the five reported instances, the positive $\delta^{84}\text{Sr}_{86}$ anomaly remains at $>2\sigma$ even if the background corrections were artificially increased by 10%.

### A.5 Subgrain Detection

RIMS data were collected over a period of tens of minutes, at a rate of 1000 laser shots per second. We binned the data into 30–100 s packets (30,000 to 100,000 laser shots) to look for signal bursts that could indicate Sr-rich subgrains. Figure A2 illustrates one such analysis (grain KFB1h-412). When such bursts were located, we binned them into smaller packets to isolate the bursts from the surrounding low-level Sr signal. We then checked these bursts for dead-time problems. A positive $\delta^{84}\text{Sr}_{86}$ could result from undercounting $^{86}$Sr if the arrival rate of $^{86}$Sr is high enough that some ions hit the detector during its dead time while it is rearming itself after an ion strike (~1.5 ns for our system). Because the abundance of

$^{86}$Sr is ~20 times higher than $^{84}$Sr, this could result in the detector missing $^{86}$Sr ions but not $^{84}$Sr and thus produce an artificially high $\delta^{84}$Sr$_{86}$. We checked for dead-time effects by tracking $\delta^{88}$Sr$_{86}$. Because $\delta^{84}$Sr$_{86}$ is close to zero in all of our grains and subgrains, and $^{88}$Sr is ~8 times more abundant than $^{86}$Sr, a significantly negative $\delta^{88}$Sr$_{86}$ during a signal burst indicates a likely dead-time issue, and so all data for the burst were discarded. Using this process, we discovered that enhanced $\delta^{84}$Sr$_{86}$ came only in brief bursts, when count rates for all Sr isotopes were much higher than baseline.

Figure A2(a) shows the time evolution of the Sr and Mo signals in a grain with enhanced $\delta^{84}$Sr$_{86}$ (KFB1h-412). Strontium bursts are apparent at ~14 and ~20 minutes. There are several Mo bursts, but none coincide with Sr. We interpret this behavior as indicative of distinct Sr- and Mo-rich subgrains being exposed as the laser evaporates material from the grain over the course of the analysis. Even during the signal bursts, the large majority of the sampled material is the graphite matrix because subgrains are generally a few tens of nanometers in diameter (T. K. Croat et al. 2005) and the laser beam size is ~1–2 μm. Thus, the signal is a mixture of subgrain and matrix; however, it is apparent from Figure A2(a) that the Sr and Mo concentration in subgrains is far higher than in the matrix, such that the subgrain dominates. Figure A2(b) shows that the $\delta^{84}$Sr$_{86}$ enhancement is correlated with the signal bursts, while $\delta^{92}$Mo$_{96}$ is near zero throughout the analysis. This further suggests that the observed elemental signals are coming from different subgrains, which may originate from different regions of the star.

### A.6. AGB Models Used for Comparison with Grain Data

We compare grain data to two sets of models of stellar nucleosynthesis in AGB stars: (i) the FUNS model (D. Vescovi et al. 2020) and (ii) the Monash models (A. I. Karakas et al. 2018; M. Lugaro et al. 2018). Carbon-13 pockets, which are the major contributors of neutrons for the s-process in AGB stars, are formed in the top layer of the He intershell. The reaction, $^{12}$C(p, γ)$^{13}$N(β$^+$)$^{13}$C, occurs when part of the H envelope mixes with the abundant $^{12}$C in the He intershell to form $^{13}$C pockets (M. Lugaro 2005). The FUNS model uses magnetic buoyancy to mix materials between the H envelope and the He intershell (D. Vescovi et al. 2020; N. Liu et al. 2022), in contrast to convective overshooting that was implemented in earlier FRUITY versions (e.g., S. Cristallo et al. 2011). The Monash model uses "exponential" convective mixing for the efficient generation of $^{13}$C pockets (C. Travaglio et al. 2018). Despite the different mechanisms used by the two models, they both produce very similar stellar yields over the mass and metallicity ranges used in this study: we used 2 $M_\odot$ and $Z$ = 0.008–0.017 from the FUNS models and 2 $M_\odot$, $Z$ = 0.014, and 2.5 $M_\odot$, $Z$ = 0.03 from the Monash models.

### A.7. CCSNe and the Two-layer Mixing Model

We use postexplosion stellar isotopic abundance profiles as a function of stellar mass, for models of initial masses of 15, 20, and 25 $M_\odot$ from T. Rauscher et al. (2002) and from the NuGrid stellar library I (C. Ritter et al. 2018). Each stellar model ejecta is divided into thousands of layers, which are then classified into six to seven zones (B. S. Meyer et al. 1995) based on their one or two most abundant element(s) by mass fraction (Figure 2). We consider all ejecta layers from the base of the O/Si zone to the top of the stellar envelope, irrespective of their zones. The total number of layers and the location in mass of the base of the O/Si zones vary across stellar models, e.g., for RIT15, the O/Si zone begins at layer 173 (where 0 is the center of the star) out of a total of 2158, while for RAU15, the O/Si zone begins at layer 46 out of 653. We simulate two-layer mixing using the abundance profiles from the aforementioned models (I. Pal & H. Bouillion 2025). The goal is to verify if the mixing of products from different layers of the nucleosynthesis models (T. Rauscher et al. 2002; C. Ritter et al. 2018) can replicate the isotopic ratios measured in the grains with $^{84}$Sr excesses.

Isotopic abundances, including their radioactive precursors, were used to calculate the isotopic ratios of interest. To find the best mixing solutions constrained to the measured $\delta(^{84}$Sr/$^{86}$Sr) of the subgrains, layers with $^{84}$Sr/$^{86}$Sr greater (H) than that of a subgrain are mixed with layers of lower ratios (L) to reproduce the relative $^{84}$Sr abundance of that subgrain, according to the following:

$$G = \frac{k \times {}^{84}\text{Sr}_H + (1-k) \times {}^{84}\text{Sr}_L}{k \times {}^{86}\text{Sr}_H + (1-k) \times {}^{86}\text{Sr}_L}, \quad (A2)$$

where $k$ is the mixing factor, which ranges from 0 to 1, and $G$ stands for the $^{84}$Sr/$^{86}$Sr isotopic ratios of the grains, while $^{84}$Sr and $^{86}$Sr are the abundances in the stellar ejecta layers. For each mixing combination, the solution for $k$ was used to calculate the resultant $^{87}$Sr/$^{86}$Sr and $^{88}$Sr/$^{86}$Sr ratios, as well as the least-square difference between the measured Sr ratios and the mixed model composition. Their C/O ratios are also found from the mixing factor and the end-member abundances, and ratios less than 1 were excluded. For each subgrain, the mixing combination with the lowest residual is normalized to solar values and expressed as $\delta$-values in Figure 3. Table A1 lists the exact, model-specific, single-layer numbers used in the mixing calculations presented in Figure 3.

**Table A1**
The Model-specific Layer Numbers Used in the Mixing Calculations (Figure 3) and the Corresponding Proportions That Were Mixed to Reproduce the Sr Isotopic Compositions in the Subgrains

| Subgrain | RAU15 | | RIT15 | | RIT20 | |
|---|---|---|---|---|---|---|
| | Model layer# in O/Si zone (mixing %) | Model layer# in He/C zone (mixing %) | Model layer# in O/Si zone (mixing %) | Model layer# in He/C zone (mixing %) | Model layer# in O/Si zone (mixing %) | Model layer# in He/C zone (mixing %) |
| KFB1h-011 | 58 (1.02%) | 262 (98.98%) | 212 (0.38%) | 1373 (99.62%) | 15 (0.37%) | 1241 (99.63%) |
| KFB1h-035 | 55 (4.08%) | 264 (95.92%) | 208 (2.34%) | 1372 (97.66%) | 14 (1.26%) | 1241 (98.74%) |
| KFB1h-241 | 55 (3.36%) | 264 (96.64%) | 206 (4.10%) | 1373 (95.90%) | 21 (0.23%) | 1241 (99.77%) |
| KFB1h-412 | 56 (0.91%) | 264 (99.09%) | 204 (7.05%) | 1374 (92.95%) | 34 (0.06%) | 1241 (99.94%) |
| KFB1h-552 | 55 (2.58%) | 264 (97.42%) | 212 (0.49%) | 1372 (99.51%) | 13 (1.28%) | 1240 (98.72%) |

RAU15, RIT15, and RIT20 were able to recreate the Sr isotopic variation in the target subgrains (Figure 3). The two regions in the stellar models where the best-possible mixing layers come from are highlighted with purple shaded regions in Figure 2. We were unable to reproduce positive $\delta^{84}Sr_{86}$ through two-layer mixing using RIT20, RIT25, and RAU25, as none of the mixing solutions had C/O > 1.

We mixed the isotopes of Mo in the same proportions for each of the subgrains, and that resulted in excesses of $\delta(^{92}Mo/^{96}Mo)$, which is expected as $^{92}Mo$ is also a *p*-nuclide. We did not observe any *p*-nuclide excesses in Mo, and we attribute this as indirect evidence to the presence of different types of subgrains with different origins and to the terrestrial contamination of the grain matrix. See Section 2.1 for a discussion of contamination in the grains.

The code for the mixing model by I. Pal & H. Bouillion (2025) is available at doi:10.5281/zenodo.17027664.

# Appendix B
## Justification for Using the C/O > 1 Constraint in Mixing Calculations

Several studies explain the production of C dust in O-rich SN environments (D. D. Clayton et al. 1999, 2001, D. D. Clayton 2011, E. A. N. Deneault et al. 2006), and this is a matter of debate in the community. However, there is no isotopic (or microstructural) evidence in C-rich presolar grains that supports this scenario. O-rich SN environments are known to be highly $^{12}C$- and $^{16}O$-rich, whereas SN graphite grains studied so far have a wide range of $^{12}C/^{13}C$ ratios and are highly enriched in $^{15}N$ and $^{18}O$ (e.g., M. Jadhav et al. 2013b; S. Amari et al. 2014). Furthermore, microstructural studies by transmission electron microscopy (TEM) of presolar graphite grains have only found reduced subgrain phases within graphites—never any oxide subgrains (T. J. Bernatowicz & R. Cowsik 1997). All the subgrain phases discovered within graphites provide tight constraints to the pressure and temperature conditions at the time of condensation and more importantly, the C/O ratio of the gas. No phases indicate that the C/O ratio should be less than 1 (T. J. Bernatowicz & R. Cowsik 1997; T. K. Croat et al. 2003, 2005). Finally, the source of the $^{84}Sr$ anomalies in the present study are Mo-, Zr-, and Sr- rich subgrains; these subgrains had to have condensed prior to the host graphite and TEM studies in conjunction with equilibrium thermodynamics condensation sequence calculations have shown that this happens only in C-rich environments (T. J. Bernatowicz et al. 1996, T. J. Bernatowicz & R. Cowsik 1997; T. K. Croat et al. 2003, 2005). Thus, in the absence of evidence to the contrary, we use the C/O > 1 constraint for the mixing calculations.

# Appendix C
## Could the Grains Originate in an AGB Star That Formed from a Molecular Cloud Already Enriched in $^{84}Sr$?

The abundance of $^{86}Sr$ increases by a factor of ~10 in AGB star envelopes, indicating that no matter the initial abundance of $^{84}Sr$, the value of $\delta^{84}Sr_{86}$ will always be ~−1000‰. The value of $\delta^{84}Sr_{86}$ is controlled by the abundance of $^{86}Sr$ and not by that of $^{84}Sr$, and thus, the anomalous grains in this study cannot originate from AGB stars that were formed from a molecular cloud that was already enriched in $^{84}Sr$. This is demonstrated by an example in Figure C1. The final composition of the stellar envelope of a low-metallicity AGB star (Monash model: A. I. Karakas et al. 2018; M. Lugaro et al. 2018) starting with twice the amount of $^{84}Sr$ is essentially identical to that of the ones with 1X abundance of $^{84}Sr$.

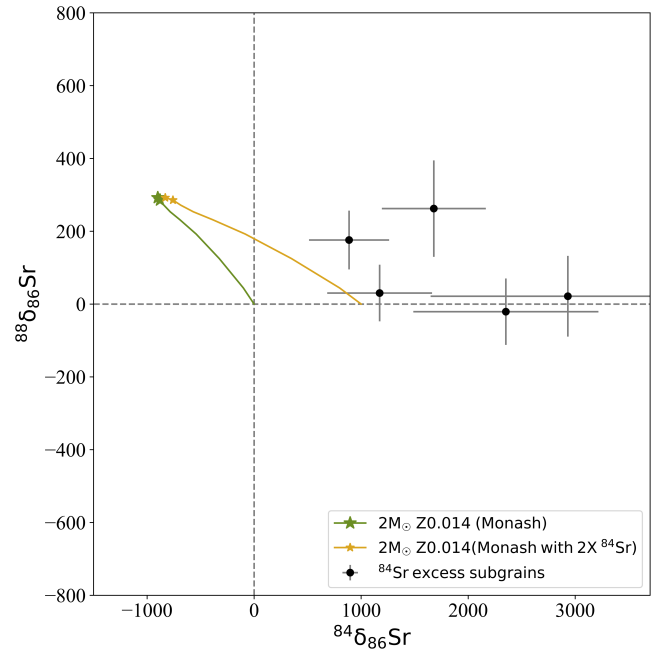

**Figure C1.** Three-isotope Sr plots of presolar HD graphites from KFB1 Murchison analyzed in this work. The five subgrains with positive $\delta^{84}Sr_{86}$ values are indicated by solid black circles. All the uncertainties represent 1σ variation. Isotopic ratios of the envelopes in low-metallicity AGB stars (A. I. Karakas et al. 2018; M. Lugaro et al. 2018) with 1X and 2X initial abundances of $^{84}Sr$ are shown as green and yellow colored lines, respectively. Markers along the prediction lines indicate isotopic ratios when the stellar envelope is C rich. These yellow and green markers have almost identical isotopic ratios, indicating that there is essentially no difference in the abundance of $^{84}Sr$ for the 1X and 2X cases in C-rich envelopes of low-metallicity AGB stars.

# Appendix D
## Two Alternative Explanations for $^{84}Sr$ Excesses from Massive Stars Measured in Presolar Graphite Subgrains

### D.1. Alternative Explanation 1: Anomalous Pristine Sr Composition of Massive Star Progenitors of Graphites

Here we explore an alternative scenario that can reproduce the measured $^{84}Sr$ excesses by using only the outer layers of the CCSN ejecta. Instead of mixing the C-rich He/C zone with the internal O/Si zone (as discussed in Section 2.2), we consider mixing with the outer stellar envelope. We then varied the initial abundances of $^{84}Sr$ and $^{86}Sr$ in the nucleosynthesis models (T. Rauscher et al. 2002; C. Ritter et al. 2018) to investigate how they affect the resultant nucleosynthetic yield of $^{84}Sr$ in the He shell of each model. We altered the initial Sr abundances in three ways: (i) reduced initial $^{86}Sr$ abundance by a factor of 2, (ii) increased initial $^{84}Sr$ abundance by a factor of 2, and (iii) increased initial $^{84}Sr$ abundance by a factor of 3. For simplicity, we assume that the $^{84}Sr$ consumed by *s*-process in the He shell remains the same, despite changing the initial amount of $^{84}Sr$. The isotopic abundance evaluated for the surface of the star is used as the initial reference value because the surface composition remains largely unmodified during the evolution of the stellar model. Sr abundances from the He shell are taken from a representative layer of the He/C zone. The difference between the surface and the He-shell value of the unaltered model is assumed to be the $^{84}$Sr consumed by the $s$-process. We observed that when the initial $^{84}$Sr is increased by a factor of 3, the resultant $\delta^{84}$Sr$_{86}$ of the He shell is positive and within the range of the $^{84}$Sr excesses observed in the presolar graphite subgrains of this study. In RIT15, the $\delta^{84}$Sr$_{86}$ value becomes positive ($\delta^{84}$Sr$_{86}$ ~ 5‰) with only a twofold increase of initial $^{84}$Sr.

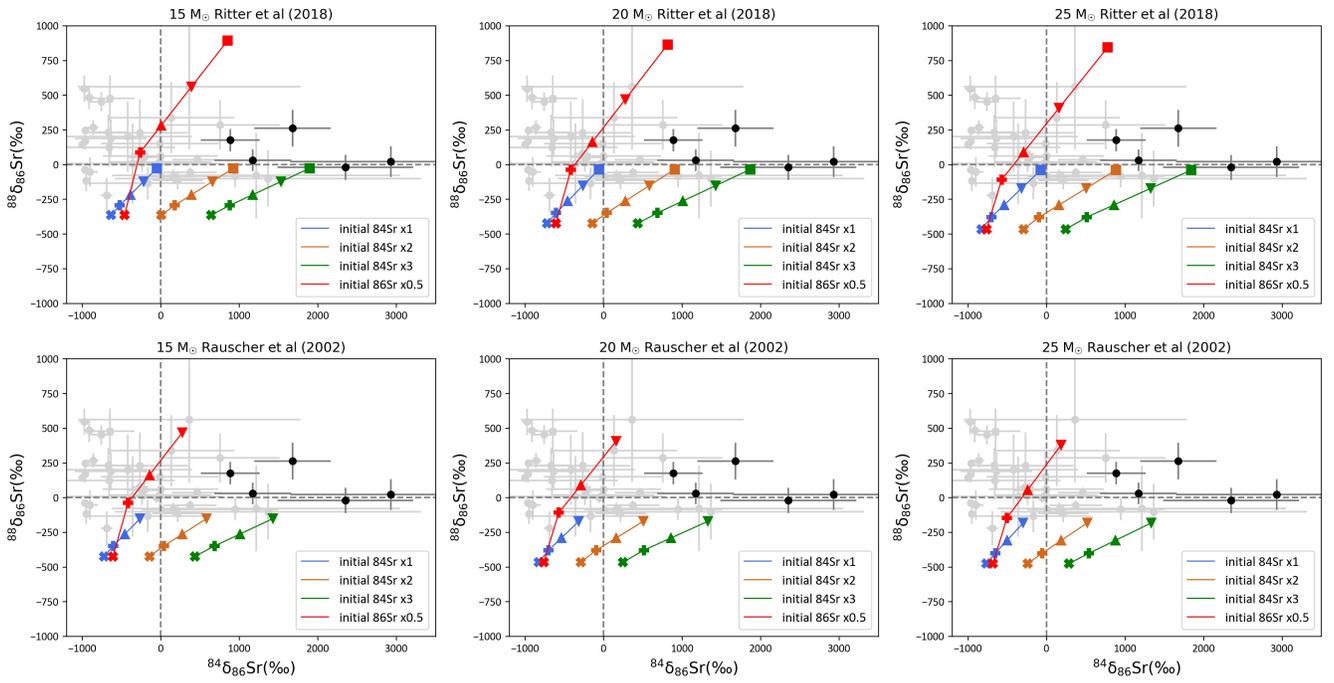

**Figure D1.** An alternative explanation for $^{84}$Sr enrichments in our graphite grains. The figure shows six $\delta^{88}$Sr$_{86}$ versus $\delta^{84}$Sr$_{86}$ subplots with measured data on 49 HD graphite grains from the KFB1 fraction of Murchison. The five subgrains with positive $\delta^{84}$Sr$_{86}$ values are indicated in black. The subplots are overlaid with the resultant isotopic values from varying initial Sr compositions and dilution of the He shell by the envelopes of the model stars. Each subplot shows the calculated results from a different stellar model: RIT15, RIT20, RIT25, RAU25, RAU20, and RAU15, respectively (moving clockwise from the upper left corner). The colored lines in each subplot represent different initial Sr compositions of the respective model star (see legend). In each colored line, different marker symbols represent different percentages of contribution of the envelope relative to the He shell: the "x" sign (0%), the plus sign (25%), upright triangle (50%), inverted triangle (75%), and square box (95%). Only the combinations with C/O > 0.99 are plotted. All uncertainties shown are 1$\sigma$.

We further explore the scenario where the He-shell output is diluted with varying proportions of material from the envelope of the star. For each of the altered cases, we consider three different mixing cases between the He shell and the envelope of the star. We calculate $\delta$-values by mixing 25%, 50%, 75%, and 95% of envelope abundances with that of the He shell (Figure D1). For the original solar metallicity model of $Z = 0.02$ (C. Ritter et al. 2018), we observe that no amount of mixing with the envelope can produce positive delta values. A doubled $^{84}$Sr initial enhancement requires 50% envelope material to be mixed to reproduce the positive $\delta^{84}$Sr$_{86}$ values measured in the subgrains. Increasing the initial $^{84}$Sr by 3 times easily produces positive $\delta^{84}$Sr$_{86}$ without any dilution. Reducing the initial $^{86}$Sr abundances instead requires 75% dilution by envelope material to obtain a significant, positive $\delta^{84}$Sr$_{86}$. Further dilution lowers the C/O ratio to less than 1, which is not conducive to graphite condensation. Note that the enhancement of initial $^{84}$Sr by a factor of 2 with 50% dilution produces the same $^{84}$Sr as reduction of $^{86}$Sr by a factor of 2 with 75% dilution by the envelope abundances.

Therefore, we find that if the initial $^{84}$Sr is enhanced, a 50% mixing with the envelope can reproduce the positive $\delta^{84}$Sr$_{86}$ values measured in the grains. Increasing the initial $^{84}$Sr by 3 times easily produces positive $\delta^{84}$Sr$_{86}$ without any dilution. Instead, for reduced initial $^{86}$Sr abundance, at least a 75% dilution is needed to reproduce the observed positive $\delta^{84}$Sr$_{86}$. Further dilution lowers the C/O ratio to less than 1, which is not the appropriate chemical environment for graphite condensation.

Thus, we show that if the initial $^{84}$Sr abundance in the parent star is enhanced by a factor of 2 or more, such a star could produce grains that have $^{84}$Sr excesses that are inherited from Galactic chemical evolution and not due to intrinsic $p$-process nucleosynthesis in the parent star (Figure D1). The positive $\delta^{84}$Sr$_{86}$ values would be obtained from a mixture of the He-shell region of the star and the normal stellar component at the surface. Such a scenario would provide a good fit for the isotopic data measured in the subgrains found in HD graphites, without invoking extensive and tuned mixing between layers from different parts of the CCSN ejecta. However, more than 98% of the widely studied mainstream SiC grains show isotopic values for $p$-process nuclei of Mo (and for the few available data also for Ba) that are consistent with the solar system (T. Stephan et al. 2019). Therefore, based on current observations, an initial Sr isotopic composition much different than solar for the parent stars seems to be unlikely. This would require that the group of HD graphites and their subgrains used in this study have a very anomalous progenesis, not represented by the majority of the presolar dust population. Therefore, at present, we consider the more likely scenario presented in Section 3, where the positive $\delta^{84}$Sr$_{86}$ values are signatures of the $\gamma$-process instead of being a result of Galactic chemical evolution at the time of the formation of the Sun in the Milky Way disk.

*D.2. Alternative Explanation 2: Thermonuclear Supernovae as a Source of Presolar Graphite grains*

Another potentially relevant scenario for this work that has been proposed in the literature is that C-rich dust can also form in the outer regions of exploding white dwarf stars or SNe Ia (D. D. Clayton et al. 1997). Also, theoretical stellar simulations studies have shown that *p*-process nuclei can be produced in the external ejecta of Chandrasekhar-mass SNe Ia (C. Travaglio et al. 2011, 2018; M. Kusakabe et al. 2011; U. Battino et al. 2020). Thus, we considered SNe Ia as alternative progenitors for the presolar graphite grains with *p*-process anomalies discussed in this study. We concluded that this scenario is unlikely for the following reasons: (1) No isotopic data on C-rich presolar grains show any evidence of having condensed in SN Ia outflows (E. Zinner 2014; L. R. Nittler & F. Ciesla 2016). (2) There has been no astronomical detection of newly formed dust in the ejecta of SN Ia remnants (H. L. Gomez et al. 2012), and the lack of this evidence is consistent with predictions from theoretical simulations by T. Nozawa et al. (2011). A very recent study reported the first observation of some dust formation 3 yr after an SN Ia explosion in the cold dense region behind the ejecta and circumstellar medium (CSM) interaction (Q. Wang et al. 2024). However, such interactions between SN Ia explosions and CSM are rare (J. M. Silverman et al. 2013), and it is unclear what the composition of this dust would be and whether it would carry any nucleosynthetic signatures from the explosion. (3) Predictions of *p*-process nuclei production have only been obtained from Chandrasekhar-mass SNe Ia that evolved within the single-degenerate scenario (e.g., T. K. Croat et al. 2003 and references therein). The evolution channel leading to SNe Ia is a matter of debate in the literature, but there are theoretical and observational indications that the single-degenerate channel evolution pathway relevant here may be a rare source of SNe Ia explosions (J. Johansson et al. 2016; P. A. Denissenkov et al. 2017; U. Battino et al. 2020). Thus, in the absence of compelling evidence for other stellar sources, we reiterate that the CCSN explanation is the most likely explanation for the *p*-process enhancements observed in the grains in this study.

The future discovery of rare types of presolar grains with evidence for SN Ia origins would require us to reconsider the astrophysical scenario discussed in the present study and would indeed present novel opportunities for cosmochemistry studies.

## ORCID iDs

Ishita Pal ● https://orcid.org/0009-0004-9537-1425
Manavi Jadhav ● https://orcid.org/0000-0002-2859-9137
Danielle Z. Shulaker ● https://orcid.org/0000-0001-8028-8670
Michael R. Savina ● https://orcid.org/0000-0002-3077-618X
Marco Pignatari ● https://orcid.org/0000-0002-9048-6010
Lorenzo Roberti ● https://orcid.org/0000-0003-0390-8770
Heather Bouillion ● https://orcid.org/0009-0008-5618-7368
Maria Lugaro ● https://orcid.org/0000-0002-6972-3958
Noriko Kita ● https://orcid.org/0000-0002-0204-0765
Sachiko Amari ● https://orcid.org/0000-0003-4899-0974